\documentclass[pra,aps,twocolumn,10pt,superscriptaddress,notitlepage,longbibliography]{revtex4-2}

\usepackage[dvipsnames]{xcolor}
\usepackage[english]{babel} 
\usepackage{graphicx}
\usepackage{float}
\usepackage{braket}
\usepackage{epstopdf}
\usepackage{mathtools}
\usepackage{amsmath}
\usepackage{amssymb}
\usepackage{bbm,bm}
\usepackage[caption=false]{subfig}
\usepackage{pythonhighlight}
\usepackage{hyperref}
\usepackage{float}
\usepackage{bbold}
\usepackage[T1]{fontenc}

\usepackage[markup=underlined]{changes}
\makeatletter
\@namedef{Changes@AuthorColor}{magenta}
\colorlet{Changes@Color}{magenta}
\makeatother
\usepackage{hyperref}
 \hypersetup{
     colorlinks=true,
     linkcolor=blue,
     filecolor=blue,
     citecolor = magenta,      
     urlcolor=red,
     }

\usepackage{braket}

\definecolor{cinnamon}{rgb}{0.82, 0.41, 0.12}
\definecolor{brown}{rgb}{0.65, 0.16, 0.16}

\usepackage{xargs}
\newcommandx{\greencom}[2][1=]
{\todo[inline, color=green!40,#1]{#2}}
\newcommandx{\bluecom}[2][1=]
{\todo[inline, color=blue!40,#1]{#2}}
\newcommandx{\bluemargin}[2][1=]
{\todo[color=blue!40,#1]{#2}}

\usepackage{letltxmacro}
\LetLtxMacro{\ORIGselectlanguage}{\selectlanguage}
\makeatletter
\DeclareRobustCommand{\selectlanguage}[1]{%
  \@ifundefined{alias@\string#1}
    {\ORIGselectlanguage{#1}}
    {\begingroup\edef\x{\endgroup
       \noexpand\ORIGselectlanguage{\@nameuse{alias@#1}}}\x}%
}
\newcommand{\definelanguagealias}[2]{%
  \@namedef{alias@#1}{#2}%
}

\makeatother

\definelanguagealias{en}{english}
\definelanguagealias{EN}{english}
\definelanguagealias{eng}{english}
\definelanguagealias{de}{ngerman}

\graphicspath{{paper_figs/}}

\begin{document}

\title{Matrix Product State Theory of Few-Photon Squeezed Pulses Interacting with a Two-Level Emitter in a Waveguide
}

\author{Sofia Arranz Regidor}
\email{18sar4@queensu.ca}
\affiliation{Department of Physics,
Engineering Physics and Astronomy, Queen's University, Kingston, Ontario, Canada, K7L 3N6}
\author{Matthew Kozma}
\email{24nkr1@queensu.ca}
\affiliation{Department of Physics,
Engineering Physics and Astronomy, Queen's University, Kingston, Ontario, Canada, K7L 3N6}
\author{Stephen Hughes}
\email{shughes@queensu.ca}
\affiliation{Department of Physics,
Engineering Physics and Astronomy, Queen's University, Kingston, Ontario, Canada, K7L 3N6}

\date{\today}

\begin{abstract} 
Squeezed light states have a special place in quantum optics with potentially profound applications in emerging quantum technologies. We present a 
numerically-exact matrix product states (MPS) approach to model
quantum pulses of squeezed light, at the few-photon level, interacting with a two-level system in a waveguide environment. 
We represent the squeezed state as a coherent superposition of Fock states and explore the nonlinear population dynamics as well as multi-photon correlation functions that emerge. We show how the squeezed pulse can create quantum correlations that are unique to squeezed pulses, including
$\braket{b(t) b(t+t')}$ for transmitted fields as well as
$\braket{b^\dagger(t) b^\dagger(t+t') b(t+t')
b(t)}$. 
We also demonstrate how $\braket{b(t) b(t+t')}$, a first-order correlation function, shows nonlinear photon correlations that are similar to those known and measured for 
two-photon scattering states.
Finally, we also study the squeezed spectra of the pulse before and after interacting with the two-level system, and highlight the role of the spectral bandwidth of the incident pulse. The MPS theory allows the modeling of arbitrary bandwidth squeezing without making any Markov and Born approximations for the light-matter interaction processes, and can easily be extended to waveguide systems with multiple emitters and
time-delayed feedback. 

\end{abstract}

\maketitle

\section{Introduction}
\label{sec:intro}

Squeezed light is one of the most fundamental nonclassical states of the electromagnetic field---a quantum state of light that is characterized by a reduction of the uncertainty in one notable property, such as amplitude or phase, below the standard quantum limit~\cite{Andersen_2016,SCHNABEL20171,Walls1983}.
Since it is one of the simplest manifestations of quantum optical behavior, it has been deeply studied in nonlinear and quantum optics, playing an important role in a wide range of quantum applications in quantum technologies. These include quantum-enhanced sensing and metrology, quantum LIDAR, quantum information processing and ultra-low noise precision measurements, and quantum computing~\cite{Lu2023}. A well-known example of quantum-enhanced sensing is achieved by using squeezed light for LIGO, improving the sensitivity of gravitational wave detectors~\cite{Caves1981,Aasi2013,PhysRevLett.123.231107,Acernese2019,Barsotti2019}.  

For several decades, experiments on squeezed light generation have greatly advanced since the early experiments using nonlinear optical processes~\cite{Slusher1985,PhysRevLett.57.2520,Vahlbruch2008,Vahlbruch2016}. 
Optical parametric processes have enabled progressively stronger squeezing, including experimental demonstrations reaching the 10-dB regime~\cite{PhysRevLett.100.033602}
and 15-dB regime~\cite{Vahlbruch2016}, with ongoing
improvements. In addition, squeezed light generation is not limited to optical and photonic systems, and, for example, similar states can be generated and controlled in superconducting microwave circuits~\cite{Yurke1988,Movshovich1990,CastellanosBeltran2008,Mallet2011,Eichler2011,Macklin2015,PhysRevLett.101.253602}.

Although squeezed light experiments are well-established in different platforms, the role of squeezed light as a fundamental quantum input for nonlinear light-matter interactions, particularly when confined to 
finite-duration pulses {\it containing only a few photons}, remains less explored.

In studies with large photon numbers, a quantum operator is well described in terms of a mean value plus quantum fluctuations, and the common properties to connect to are classical-like amplitude and phase. In contrast, in the few-photon regime, quantum fluctuations can dominate the dynamics, and an exact Fock state basis is ultimately needed when exploring such quantum correlations. 
Yet squeezed states are substantially different to 
Fock states, especially with regard to photon-photon correlations.
At a fundamental level, we expect the interaction of many-photon squeezed states and few-photon states to behave differently. Despite this fundamental difference, 
theoretical research has been mainly done in the continuous wave (CW) limit, where the squeezing is constant over all frequencies.
Relatively few works have explored the fundamentals of few-photon squeezed light pulses, and their interaction with a two-level system (TLS)~\cite{PhysRevA.105.023721}. From a theoretical perspective, this is also largely due to the potential complexity of treating such fields exactly.

In contrast, there has been an increasing number of studies exploring how few photons, and often just two photons, interact with 
a two-level system (TLS) in a waveguide. This is often in the regime of waveguide-QED, which can be realized in a number of practical configurations, including atoms in fibers~\cite{PhysRevA.50.2680,PhysRevLett.104.203603}, quantum dots in semiconductor waveguides~\cite{PhysRevB.99.085311,PhysRevB.75.205437,PhysRevLett.113.093603,Paesani2019,leFeber2015,PhysRevLett.115.153901,Sllner2015,doi:10.1126/sciadv.aaw0297,PhysRevResearch.4.023082,Liu:22,PhysRevLett.117.240501,PhysRevX.10.031011}, and circuit-QED systems~\cite{RevModPhys.93.025005,2020blais,QIN20241}. 
One of the most striking examples is a strong two-photon nonlinearity as well as a pronounced population transfer that can occur from two-photon Fock states. Two-photon scattering is often described using
scattering matrices and input-output theory~\cite{PhysRevA.76.062709,PhysRevLett.98.153003,PhysRevLett.126.023603},
but is difficult to extend to higher quanta (e.g., a higher number of photons or more TLSs).
In contrast, matrix product states (MPS) theory is an efficient and numerically exact way (bound by truncation, which can easily be checked) of representing an arbitrary number of photons and quantum emitters. Although some works have shown how to model Fock-state pulses with MPS~\cite{PhysRevA.103.033704,lp1b-yswm}, 
to our knowledge, we are not aware of any works showing how to model squeezed
few-photon Fock pulses with either scattering theory or MPS. 

In this work, we develop an MPS framework to study finite-duration squeezed vacuum pulses interacting with a TLS in a one-dimensional waveguide. The approach represents a squeezed input pulse directly within the discretized waveguide Hilbert space, allowing its complete temporal envelope to be considered throughout the dynamics. 
This provides a numerically controlled description of squeezed-pulse scattering beyond the CW limit and complements a recent master equation approach developed for squeezed wave packets~\cite{PhysRevA.105.023721}. Our MPS formalism gives us full access to field observables, allowing us to calculate field correlations and output spectra in a straightforward way. 
More broadly, our method provides a route for studying arbitrary pulsed squeezed states in waveguide-QED systems in the non-Markovian regime, e.g., with delayed time feedback effects, using tensor-network techniques. 
Our finite-time approach also allows the creation 
of finite-bandwidth squeezing (from narrowband squeezing to broadband squeezing if one wishes), while fully accounting for
light-matter interactions without any Markov or Born approximations.


\begin{figure}[t]
\centering
    \includegraphics[width=\columnwidth]{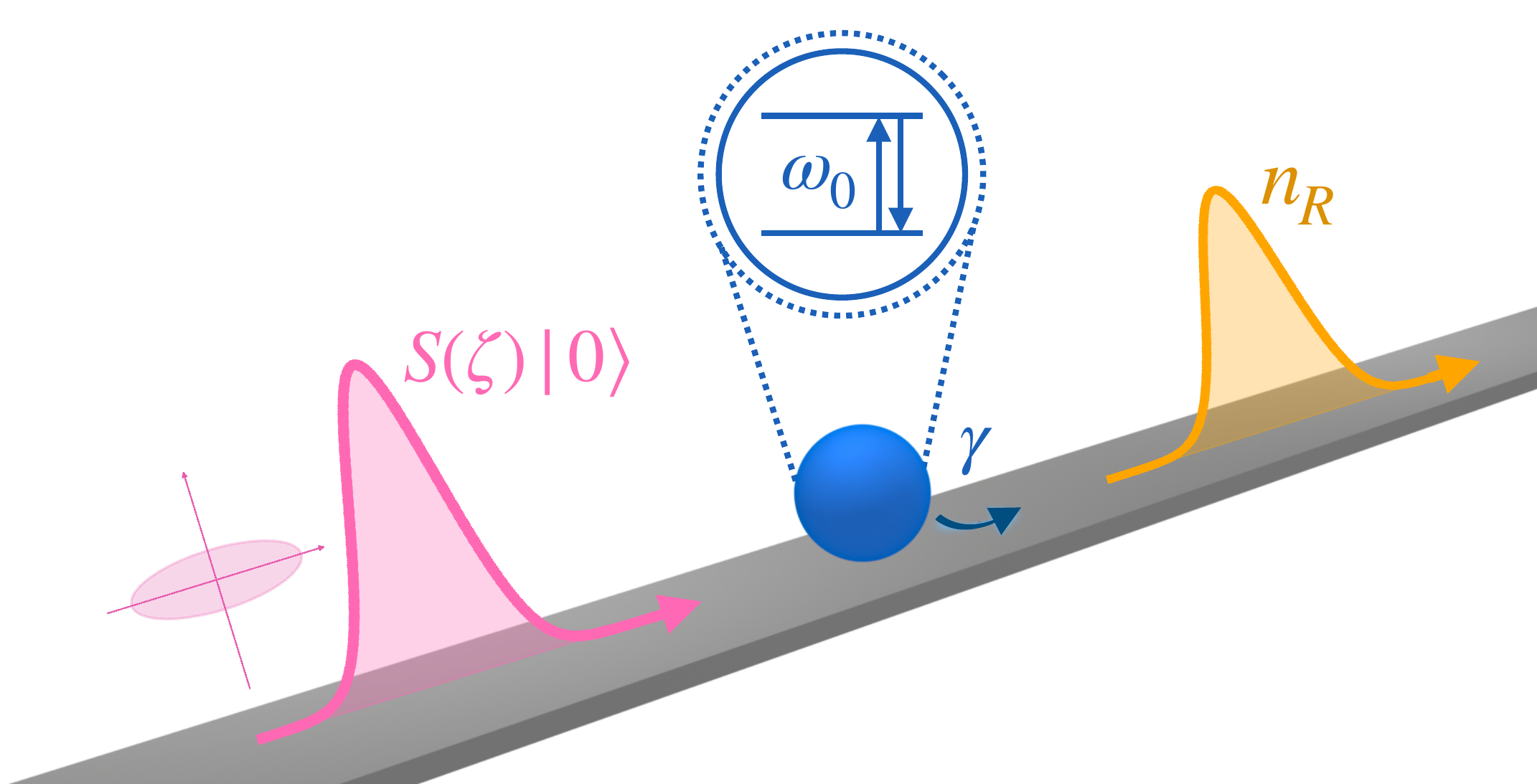}
    \caption{Schematic of an initial squeezed pulse $S(\zeta) \ket{0}$ represented in magenta interacting with a TLS (in blue) right-chirally coupled to a waveguide. The right transmitted flux $n_R$ is shown in orange (after interacting with the TLS).}
    \label{fig:schematic}
\end{figure}

The rest of our paper is organized as follows. 
In Sec.~\ref{sec:theory}, we introduce the 
waveguide-QED theory with squeezed input pulses
using MPS, and we describe how to 
create the squeezed pulses with states that are constructed
from a Fock state basis. The theory can be applied to 
any general number of photons, though it can be sensibly truncated when dealing with few-photon pulses. 
We first introduce in Sec.~\ref{subsec:hamiltonian} the Hamiltonian used to solve the problem, and in Sec.~\ref{ubsec:initial_state} we define the creation of the initial squeezed state. Local observables are introduced in Sec.~\ref{subsec:observables}, and then an extension to two-time correlation functions and spectral observables in Sec.~\ref{subsec:correlations}.
In Sec.~\ref{sec:results}, we calculate the numerical results of our system, where we start by calculating the population dynamics in Sec.~\ref{subsec:res_pop}. In Sec.~\ref{subsec:res_firstcor}, we calculate two-time first-order correlation functions and study the differences between a weak squeezed pulse and a 2-photon Fock state with a similar pulse envelope, showing correlation signatures unique to squeezed pulses. Then, we examine two-time second-order correlation functions in Sec.~\ref{subsec:res_secondcor} with a similar comparison. Section~\ref{subsec:res_spectra} gives a study of the squeezing spectra for different pulse lengths, where we show their dependence on both the bandwidth of the pulse and the nonlinear emitter interaction. 

Finally, in Sec.~\ref{sec:conclusions} we give a summary and final conclusions. In addition, in App.~\ref{appendix}, we show a numerical benchmark of the MPS population dynamics 
by comparing to an
alternative master equation solution developed for squeezed wave packets~\cite{PhysRevA.105.023721}.

\section{Modeling squeezed-state pulses with matrix product states}
\label{sec:theory}
To model our waveguide-QED system in a quantitative way (cf.~Fig.~\ref{fig:schematic}), we use MPS~\cite{PhysRevLett.116.093601}, which allows us to solve the problem without some of the common model limitations~\cite{PhysRevResearch.3.023030,Pichler11362}. For quantum circuits with delay lines and quantum feedback, the general waveguide MPS approach was introduced by Pichler and Zoller~\cite{PhysRevLett.116.093601,PhysRevA.93.062104}.
This yields a numerically exact approach, based on tensor networks (TN), that allows one to discretize the electromagnetic field into {\it time bins}, and thus, limit the growth of the Hilbert space in order to solve the problem in a numerically tractable manner~\cite{hmbj-lg7p,PhysRevX.14.031043}. 

\subsection{Hamiltonian of the system}
\label{subsec:hamiltonian}

A single TLS coupled to a waveguide can be defined with the following Hamiltonian,
\begin{equation}
    H = H_{\rm W} + H_{\rm TLS} + H_{\rm I},
\end{equation}
where
\begin{equation}
    H_{\rm W}= \sum_{\alpha=\rm L,R}\int_{\mathcal{B}} d \omega \omega b_\alpha^{ \dagger}(\omega) b_\alpha(\omega),
\end{equation}
represents the waveguide Hamiltonian, with $\alpha = \rm L,R$ the left and right channels, respectively, for the boson operators $b$, and $\mathcal{B}$ is the relevant bandwidth of interest.

The TLS Hamiltonian is
\begin{equation}
    H_{\rm TLS} =  \omega_{\rm 0} \sigma^+ \sigma^- ,
\end{equation}
with $\omega_{\rm 0}$ the TLS transition frequency,
and $\sigma^\pm$ the usual Pauli operators.

Finally, the interaction term between the TLS and the waveguide is defined, in the rotating wave approximation by,
\begin{equation}
   H_{\rm I} = \sum_{\alpha=\rm L,R}\sqrt{\frac{\gamma_\alpha}{2\pi}} \int_\mathcal{B} d \omega \left( \sigma^+ b_\alpha(\omega) + b_\alpha^{ \dagger}(\omega) \sigma^- \right),
\end{equation}
 with $\gamma_{\rm L}/\gamma_{\rm R}$ the left/right coupling rates.
 Here, we will consider a TLS on resonance with the squeezed pulse central frequency, $\omega_{\rm p}$, thus $\omega_{\rm p}=\omega_{\rm 0}$.

To write the Hamiltonian in terms of the MPS formalism, the frequency-dependent creation and annihilation operators for the waveguide can be transformed to the time domain as follows,
\begin{equation}
    b_\alpha(t)=\frac{1}{\sqrt{2\pi}} \int d\omega b_\alpha(\omega) e^{-i(\omega - \omega_{\rm 0})t},
    \label{continuous-time-ops}
\end{equation}
which correspond to the quantum noise operators with the commutation relations $\left[ 
 b_{\alpha}(t), b_{\alpha'}(t')\right] = \delta_{\alpha, \alpha'} \delta(t-t')$.
Then, they can be discretized and written in terms of the time-bin noise operators,
\begin{align}
    \label{noise1}
    &\Delta B_\alpha^{(\dagger)}(t_k) = \int_{t_k}^{t_{k+1}} dt' b_\alpha^{(\dagger)}(t')
\end{align}
where these operators create/annihilate a photon in a time bin and have the commutation relations $\left[ \Delta B_{\alpha}(t_k), \Delta B_{\alpha'}^\dagger(t_{k'}) \right] = \Delta t \delta_{k,k'} \delta_{\rm \alpha,\alpha'}$.

In this scheme, the Hamiltonian is transformed to
\begin{equation}
    H (t_k)= \sum_{\alpha={\rm L,R}} \sqrt{\frac{\gamma_\alpha}{2}} \left( \sigma^+ \Delta B_\alpha(t_k) + \sigma^- \Delta B_\alpha^{(\dagger)}(t_k) \right)/\Delta t,
\end{equation}
and then introduced in the time evolution operator, 
\begin{equation}
    U(t_k) =  \exp{ \left( -i \Delta t H(t_k)\right)},
    \label{eq:t_evol}
\end{equation}
to evolve the system at each time step. 

In the following sections, we will focus on right-scattering chiral results. Thus, we choose $\gamma_{\rm L} =0$ and $\gamma_{\rm R} =\gamma$, and we will drop the subscript $\alpha$ from boson operators for simplicity. 
This is not a model restriction, and it is straightforward to calculate similar results for a TLS symmetrically coupled to the waveguide in both directions; however, a chiral emitter system
makes the physics easier to explore and assess, and it is also of fundamental interest in recent years~\cite{PhysRevLett.115.153901,Sllner2015,leFeber2015,Lodahl2017}

\subsection {Creating the initial state and quantum pulse}
\label{ubsec:initial_state}

An initial quantum pulse in the Fock state basis, which is required for an exact few-photon representation,  can be modeled in the photonic part of the initial total state~\cite{Guimond_2017,lp1b-yswm}. Thus,  we write  an initial state as
\begin{equation}
    \ket{\psi_0} = \ket{i}_s \otimes \ket{\phi_0},
\end{equation}
where $\ket{i}_s$ corresponds to the TLS part and $\ket{\phi_0}$ to the photonic part, and each time bin contains a tensor product of the right and left channels at the corresponding time step. 

Accordingly, a one-photon pulse can be introduced from
\begin{equation}
        \ket{\phi_0} = b_{\rm in}^\dagger \ket{0} , 
\end{equation}
with
\begin{equation}
    b_{\rm in}^\dagger = \int dt \ f(t) b^\dagger(t),
    \label{bin}
\end{equation}
and $f(t)$ is normalized through $\int|f(t)|^2dt=1$. 
 
In order to model an initially squeezed pulse in the MPS framework, we can start by writing the squeezed state in the Fock basis~\cite{Christ2011,PhysRevA.73.063819}. In addition, we need to consider the quantized pulse envelope. 
The squeezing operator is defined from:
\begin{equation}
    S(\zeta) = {\rm exp} \left( \frac{\zeta}{2}  \int dt dt' K(t,t') b^\dagger(t) b^\dagger(t') - {\rm H.c.} \right),
    \label{eq:S_op}
\end{equation}
where $\zeta = r e^{i\theta}$ is the squeezing parameter,  with $r$ the amplitude of the squeezing and $\theta$ its angle, $K(t,t')$ is the pulse envelope of the pair, and these pairs are created similarly to a state with Fock input pulses~\cite{regidor2026quantumdynamicsfewphotonpulsed}. In the simplest case, we can separate $K(t,t^\prime) = f(t)f(t^\prime)$ into two identical individual photon envelopes. 

In that case, the initial state in Fock state basis is,
\begin{equation}
    S(\zeta)\ket{0} = \frac{1}{\sqrt{{\rm cosh}(r)}} \sum_{n=0}^{\infty} \frac{\sqrt{(2n)!}}{2^n n!} \left[ -e^{i \theta} {\rm tanh}(r)\right]^n \ket{2n},
    \label{squeezed_state}
\end{equation}
where  $2n \in \{0,2, ..., N\}$ with $N$ the total number of photons created in pairs that are considered. In the general case, $N \to \infty$, though in practice we can truncate this expansion depending on the pulse properties, which is especially useful for low photon-number pulses.

The last term of Eq.~\eqref{squeezed_state} can be written as,
\begin{equation}
\begin{split}
        \ket{2n} &= \frac{\left[ \sum_{k,l}  f(t_k)f(t_l)\Delta B^\dagger(t_k)\Delta B^\dagger(t_l) \right]^{n}}{\sqrt{(2n)!} } \ket{0}
    ,
\end{split}
\end{equation}
where we are now writing the time labels in their discretized version, so $t,t'$ transform to $t_k,t_l$, respectively
With this, we now write the complete squeezed state,
\begin{equation}
\begin{split}
    \ket{\psi} &= \frac{1}{\sqrt{{\rm cosh}(r)}} \sum_{n=0}^{\infty} \frac{\sqrt{(2n)!}}{2^n n!} \left[ -e^{i \theta} {\rm tanh}(r)\right]^n \\ & \times \frac{\left[ \sum_{k,l}  f(t_k)f(t_l)\Delta B^\dagger(t_k)\Delta B^\dagger(t_l) \right]^{n}}{\sqrt{(2n)!} } \ket{0}
\end{split}
\end{equation}

Note that here, we cannot simply write the state as a tensor product since the photons are being created in pairs. We need to entangle the time bins to account for the photon pairs created at different parts of the pulse. 
This is more easily seen by truncating our initial state to $n=1$, i.e., truncating it to a maximum of two photons. In this case, Eq.~\eqref{squeezed_state} reduces to,
\begin{equation}
    S(\zeta) \ket{0} =  \frac{1}{\sqrt{{\rm cosh}(r)}} \Big[ \ket{0} - \frac{\sqrt{2}}{2} e^{i \theta} {\rm tanh}(r)  \ket{2} \Big],
    \label{weak_sq}
\end{equation}
and then
\begin{equation}
\begin{split}
    \ket{\psi} 
    &=  \frac{1}{\sqrt{{\rm cosh}(r)}} \big[ \ket{0} \\
    &-\frac{e^{i \theta}}{2 } {\rm tanh}(r)\sum_{k,l}f(t_k)f(t_l)\Delta B^\dagger(t_k)\Delta B^\dagger(t_l) \ket{0}   \big] . 
    \label{eq:2phtrunc}
\end{split}
\end{equation}
%
In the second term of Eq.~\eqref{eq:2phtrunc}, we need to consider both when $k=l$ and then we populate $\ket{2}$, and when $k \neq l $, and we populate $\ket{11}$. 

In the MPS formalism, we can write the tensors corresponding to these subspaces, and store the squeezing information as the vectors $\alpha$ and $\Omega$ that later will be added to the edge tensors. If we define
\begin{equation}
    h= \frac{1}{\sqrt{{\rm cosh}(r)}},
\end{equation}
and
\begin{equation}
    g = -e^{i \theta} {\rm tanh}(r);
\end{equation}
then we can write:
\begin{equation}
    \alpha=\begin{bmatrix}
    h & 0 & 0
    \end{bmatrix},
\end{equation}
and 
\begin{equation}
    \Omega=\begin{bmatrix}
    1 \\ 0 \\ g
    \end{bmatrix}. 
\end{equation}

By defining the tensors for a single-mode squeezed state, limited to a pair of photons, as
\begin{gather}
A_k^{(0)}=\begin{bmatrix}
    1 & 0 & 0 \\
    0 & 1 & 0 \\
    0 & 0 & 1
\end{bmatrix}
,\qquad A_k^{(1)}=\begin{bmatrix}
    0 & f_k & 0 \\
    0 & 0 & f_k \\
    0 & 0 & 0
\end{bmatrix} 
, \\
\qquad A_k^{(2)}=\begin{bmatrix}
    0 & 0 & \frac{1}{\sqrt{2!}}f_k^2 \\
    0 & 0 & 0 \\
    0 & 0 & 0
\end{bmatrix}, 
\end{gather}
and by contracting the first and last tensors as $\alpha A_1[i,j]^{(l)}$ and $A_m[i,j]^{(l)}\Omega$, we get the edge tensors:
\begin{align}
A_1^{(1)}=&\begin{bmatrix}
    h & 0 & 0
\end{bmatrix},\\\nonumber
A_1^{(1)}=&\begin{bmatrix}
    0 & hf_1 & 0
\end{bmatrix},\\\nonumber
A_1^{(2)}=&\begin{bmatrix}
    0 & 0 & h\frac{f_1^2}{\sqrt{2!}}
\end{bmatrix},
\end{align}
\begin{align}
    A_m^{(0)}=\begin{bmatrix}
    1 \\ 0 \\ g
    \end{bmatrix}, \,
    A_m^{(1)}=\begin{bmatrix}
    0 \\ gf_m \\ 0
    \end{bmatrix}, \,
    A_m^{(2)}=\begin{bmatrix}
    g\frac{f_m^2}{\sqrt{2!}} \\ 0 \\ 0
    \end{bmatrix}.
\end{align}


For constructing states with a higher number of photons, we can write a more general squeezed state by following the same procedure as is done for Fock states in Ref.~\cite{regidor2026quantumdynamicsfewphotonpulsed}, and adding the squeezing information in $\alpha$ and $\Omega$, as in the two-photon subspace solution. In this case, we will have the following:
\begin{align}
    A_k^{(l)}[i,j] =& \delta_{i+l,j}\frac{f_k^l}{\sqrt{l!}},
\end{align}
for the general tensors, and the first and last tensors need to be contracted with the following $\alpha$ and $\Omega$,
\begin{equation}
    A_1^{(l)}[j] = \alpha A_1^{(l)}[i,j], 
\end{equation}
with $\alpha = \delta_{0,i} h$ and $h = \frac{1}{\sqrt{{\rm cosh}(r)}}$ as before, and
\begin{equation}
    A_m^{(l)}[j] = A_m^{(l)}[i,j] \, \Omega,  
\end{equation}
with
\begin{equation}
\Omega = \delta_{2j,n}  \frac{(2n)!}{2^n n!} g^n,
\end{equation}
where $g = -e^{i \theta} {\rm tanh}(r)$ is defined as before, and the total length of the vector is $2n$. Note here that in the numerator of $\Omega$ we have $(2n)!$ instead of $\sqrt{(2n)!}$ as in Eq.~\eqref{squeezed_state}; this is because we are dividing already in the tensors by $\sqrt{(2n)!}$; otherwise, we would be doing it twice.

\subsection{Local observables}
\label{subsec:observables}

First, we define the population dynamics used later to calculate local observables, including the following: the TLS population,
\begin{equation}
    n_{\rm TLS}(t) = \braket{\sigma^+ (t) \sigma^- (t)},
\end{equation}
the right output flux,
\begin{equation}
    n_{\rm R}(t) = \braket{b^\dagger (t) b(t)},
\end{equation}
and the total integrated flux, which in this case is
\begin{equation}
    N_{\rm total} (t) = n_{\rm TLS} (t) + \int_0^t n_{\rm R}(t^\prime) dt^\prime.
\end{equation}

Generally, there are certain observables that one can study to characterize the type of quantum state. In Table~\ref{tab:1}, we summarize the typical behavior of local observables studied for well defined coherent, Fock and squeezed vacuum states.

\begin{table}[h]
    \centering
    \begin{tabular}{c|c|c|c}
        Observable &  Coherent & Fock & Squeezed vacuum\\
        $\braket{b(t)}$  & $\neq 0$ & 0 & 0 \\
        $\braket{b^2(t)}$  & $\neq 0$ & 0 &  $\neq 0$ \\
        $\braket{b^\dagger(t) b(t)}$  & $= |\braket{b^2}|$ & $\neq 0$  &  $\neq 0$ \\
    \end{tabular}
    \caption{Summary of the features of observables for different initial states.}
    \label{tab:1}
\end{table}


It is important to note that the observables shown in Table~\ref{tab:1} are local observables only and, although they are helpful as a base reference, in this work we are working with pulses that have a spread in time, and thus we need to access the corresponding full two-time correlations to have a complete picture of the dynamics. 
This will become clearer in our calculations and results below,
where the squeezing characteristics, and all light-matter interactions, are time-dependent.


\subsection{Quantum correlation functions for fields and  the squeezing spectrum}
\label{subsec:correlations}

In order to account for effects that might not be properly captured by local observables, we need to define correlation functions and the squeezing output spectrum for pulsed light.

Generally,  we need the following first-order correlation functions:  $\braket{b^\dagger(t) b(t+t')}$, $\braket{b(t+t') b(t)}$, and $\braket{X_\phi^{\dagger}(t+t') X_\phi(t)}$, where $X_{\phi}$ is the quadrature operator, 
\begin{equation}
X_{\phi}
= \frac{1}{2}(e^{i\phi} b^\dagger + e^{-i\phi}b),
\label{eq:quadrature}
\end{equation}
and $\phi$ is the quadrature angle.

The squeezing output spectrum is defined as~\cite{Combes2017,Walls2025,Carmichael:87,Collett:87},
\begin{equation}
\begin{split}
S(\omega,\phi) &\equiv  4
{\rm Re} \int_{-\infty}^{\infty} \int_{-\infty}^{\infty} dt  dt^\prime \
e^{-i(\omega-\omega_0)t^\prime} \\
&T \braket{: X_\phi^{\delta}(t+t^\prime)
X_\phi^{\delta}(t):} 
,
\label{eq:spect}
\end{split}
\end{equation}
%
where $T$ denotes time ordering, and the colons $:$ denote normal ordering with the creation operator placed to the left of the annihilation operator. Both orderings are essential for obtaining the correct solution. The fluctuation operator is simply the quadrature operator defined similarly to Eq.~\eqref{eq:quadrature},
\begin{equation}
X^\delta_{\phi}
= \frac{1}{2}(e^{i\phi} b^\dagger_\delta + e^{-i\phi}b_\delta),
\end{equation}
but here $X_\delta = X - \braket{X}$ subtracts off the coherent part. 

However, vacuum squeezed states have no coherence
(i.e., $\braket{b}=0$) and hence, from Eq.~\eqref{eq:spect} and using the MPS formalism, we
obtain the squeezed spectrum from:

%
%
\begin{equation}
\begin{split}
&S(\omega,\phi) \equiv 
{\rm Re} 
\sum_{k,l}
e^{-i (\omega-\omega_0) t_l}\\
& T \Big[
\braket{\Delta B^\dagger(t_{k+l})\Delta B(t_k)}
+\braket{\Delta B^\dagger(t_k)\Delta B(t_{k+l})} + \\
&
e^{2i\phi}\braket{\Delta B^\dagger(t_k)\Delta B^\dagger(t_{k+l})} 
+
e^{-2i\phi}\braket{\Delta B(t_{k+l})\Delta B(t_k)}\Big],
\end{split}
\end{equation}
where $T$ again ensures the time ordering of the operators.

\section{Numerical Results}
\label{sec:results}

We will now exemplify a selection of numerical results.
For the quantum input field, we chose a squeezed pulse with a Gaussian-shaped envelope, which in the time domain is defined from
\begin{equation}
    f(t) = \frac{1}{\pi^{1/4}\sqrt{\sigma_t}} 
    \exp\left \{-\frac{(t-t_c)^2}{2\sigma_t^2}\right \},
\end{equation}
with $\gamma \sigma_t =1$, 
centered at $\gamma t_c=5$. In this case, the squeezed parameter chosen is $\zeta=0.1$.


In a CW squeezed vacuum state, the mean photon number is calculated via $ \bar{n} = {\rm sin h^2 (r)}$, with $r = |\zeta| $. In our case, $\bar{n} \equiv N_{\rm total}(t \to \infty)$. 

Thus, we choose a value of the squeezing parameter $\zeta$ here that is small enough to remain in a weak regime ($\bar{n} \approx 0.01$).
This is weak, but will still induce nonlinear interactions
through multi-photon contributions, since it includes
contributions from the Fock-state components $\ket{2n}$
with $n=1,2,\cdots$
[cf.~Eq.~\eqref{squeezed_state}].

In our simulations, 
for clarity with respect to the input-output quantum pulse features, we compare the results of an initial pulse going through the waveguide without any TLS interaction, with the same pulse but now interacting with a single TLS coupled to a waveguide.  All the photon observables are calculated immediately after the TLS location (at the same time step).
In the interacting case, the TLS is chirally coupled to the waveguide, but similar results can be calculated with a TLS symmetrically coupled, as mentioned earlier.

For notational simplicity in what follows,
we will redefine a rescaled version of
$\Delta B(t_k)$ as simply
\begin{equation}
b(t_k)  \equiv \Delta B (t_k)/\Delta t,
\end{equation}
which should not be confused with the continuous boson operator defined in Eq.~\eqref{continuous-time-ops}.
Thus, as an example if we show
$\braket{b(t)b(t+t')}$, it means we are computing
$\braket{\Delta B (t) \Delta B (t+t')}/(\Delta t)^2$
in the MPS formalism, using {\it discrete} time bins.

\begin{figure}[t]
\centering
    \includegraphics[width=\columnwidth]{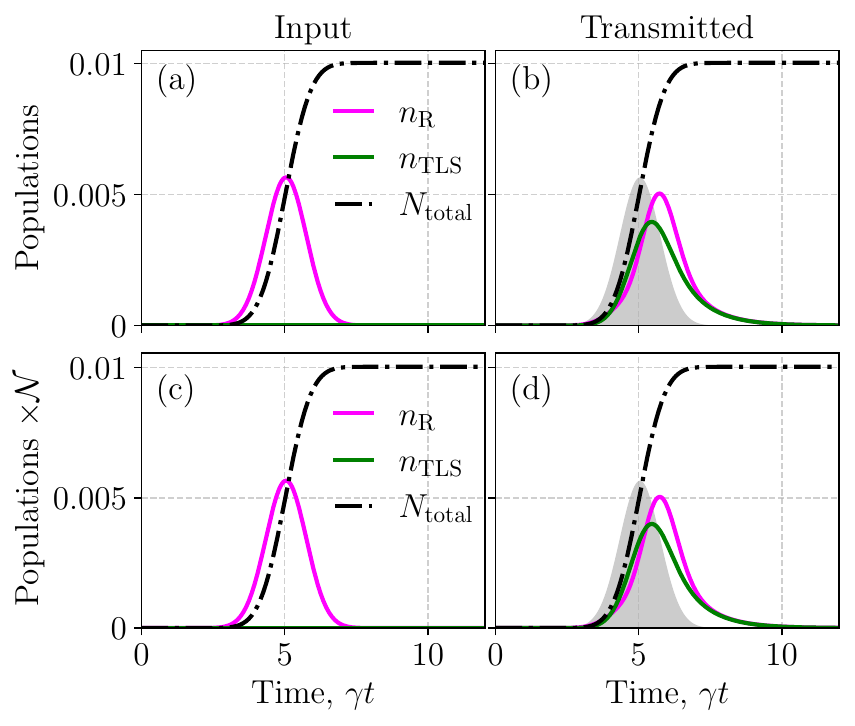}
    \caption{Population dynamics for a squeezed vacuum Gaussian pulse with $\gamma \sigma_t=1$ and centered at $\gamma t_c=5$, going through a waveguide without any interaction, labeled as input in (a) (so the emitter population remains at zero), and when interacting with a chiral TLS, labeled as transmitted, in (b). 
    For comparison, panels (c,d) show a pure 2-photon Fock Gaussian pulse normalized with $\mathcal{N}$,
    also with $\gamma \sigma_t=1$ and $\gamma t_c=5$. 
    Atomic population dynamics $n_{\rm TLS}$ is shown in green, right transmitted flux $n_{\rm R}$ in magenta, and $N_{\rm total}$ in dashed-dotted black. In (b,d), the input pulse flux is represented by the shaded grey area for reference. 
    }
    \label{fig:pop}
\end{figure}

\subsection{Population dynamics}
\label{subsec:res_pop}

In this section, we calculate the population dynamics of our system, when a pulsed few-photon 
field interacts with a (chiral) TLS in a waveguide.

For  a few-photon squeezed input state, and an initially unexcited TLS (population in the ground state), the population of the TLS, in terms of the Fock-state components, is 
\begin{align}
n_{\rm TLS}(t) &= P(0)\, n_{\rm TLS}^{N=0}(t)
+ P(2)\, n_{\rm TLS}^{N=2}(t)  
+ P(4)\, n_{\rm TLS}^{N=4}(t) \nonumber \\ 
&+ \cdots 
\nonumber \\
&\approx P(2)\, n_{\rm TLS}^{N=2}(t) \nonumber \\
&= \frac{1}{2} \frac{{\rm tanh}^2(r)}{{\rm cosh}(r)} \, n_{\rm TLS}^{N=2}(t)
\approx \frac{1}{2} {\rm sinh}^2(r) 
\, n_{\rm TLS}^{N=2}(t),
\end{align}
where we have considered that for small values of $r$, then ${\rm cosh}(r) \approx 1$ and ${\rm tanh}(r) \approx {\rm sinh}(r)$, and $P(N) = |c_N|^2$ are the probabilities of each photon subspace, with
the explicit Fock-state expansion for the input state,
\begin{equation}
\ket{\psi_{\rm in}(0)}
= \sum_N c_N \ket{N}.
\end{equation}

Thus, for a weak (low photon number) squeezed state, 
the average photon number $\bar{n} \approx P(2)$,
and the sum can be truncated to the first term of the expansion. 
In the case shown in here, $\bar{n} = {\rm sinh}^2(0.1) \approx 0.00988$ and $P(2) \approx 0.01003$. Hence, we will see that most of the results shown are dominated by the 2-photon 
subspace; however, this is not always the  case,
and the squeezed low-photon pulse also introduces
non-trivial and finite higher-order correlations.

Figure~\ref{fig:pop} shows the time dynamics for the squeezed-state pulse traveling in the waveguide (to the right) without any TLS interaction in (a), and with a TLS in (b).  The input flux is represented by the shaded grey area. The right photon flux ($n_{\rm R}$) is shown in magenta, where we can observe that when there is no interaction in Fig.~\ref{fig:pop}(a), it matches the input flux, but in Fig.~\ref{fig:pop}(b), after interacting with the TLS, it reshapes, and the peak is delayed in comparison to the reference pulse. Additionally, the TLS population ($n_{\rm TLS}$) is shown in Fig.~\ref{fig:pop}(b) in green, and the total integrated population as the pulse goes through is shown in both cases with the dashed-dotted black curve. 

Figure~\ref{fig:pop}, panels (c,d), show the time dynamics of a 2-photon Fock state with the same Gaussian envelope ($\gamma \sigma_t = 1$, $\gamma t_c=5$), again without any interaction in (c) and interacting with a TLS in (d). We observe here that the time dynamics of a weak squeezed pulse and a 2-photon Fock state are similar, apart from the overall amplitude, which is far larger in the Fock pulse solution since in this case $\bar{n}_{\rm F} = 2$ (compared to the squeezed case with $\bar{n}_{\rm sq}$). Thus, the Fock state solution is normalized with the expected values of photons in each case by defining $\mathcal{N} = \bar{n}_{\rm sq}/ \bar{n}_{\rm F} = {\rm sinh^2(r)/2}$.

Here we observe that after the interaction, when the pulse has gone through the TLS completely, $N_{\rm total}$ reaches the expected value of  ${\rm sin h}^2  (r) \approx 0.01$, both in the case without and with a TLS interaction. 

It is important to note that even for small values of $\zeta$, the number of photons necessary to consider before truncating Eq.~\eqref{squeezed_state} quickly grows. To quantify this, we calculate here a normalized deviation between the total quanta after the interaction and observe how much it deviates from 1:
\begin{equation}
    {\rm norm_{dev}} (t \to \infty) = 1- \frac{N_{\rm total}(t \to \infty)}{\rm sinh^2(r)},
    \label{eq:trunc}
\end{equation}
where $r=|\zeta|$.
For large enough $n$ in the basis expansion, then this norm is 
simply 1.

%


As an independent check for some of our numerical calculations, our MPS population dynamics results (TLS populations and transmitted photon flux) have been benchmarked against an alternative master equation solution~\cite{PhysRevA.105.023721}, shown in App.~\ref{appendix}, for three different values of the squeezed parameter, $\zeta$.

\subsection{First-order correlations}
\label{subsec:res_firstcor}

\begin{figure}[bh]
\centering
    \includegraphics[width=\columnwidth]{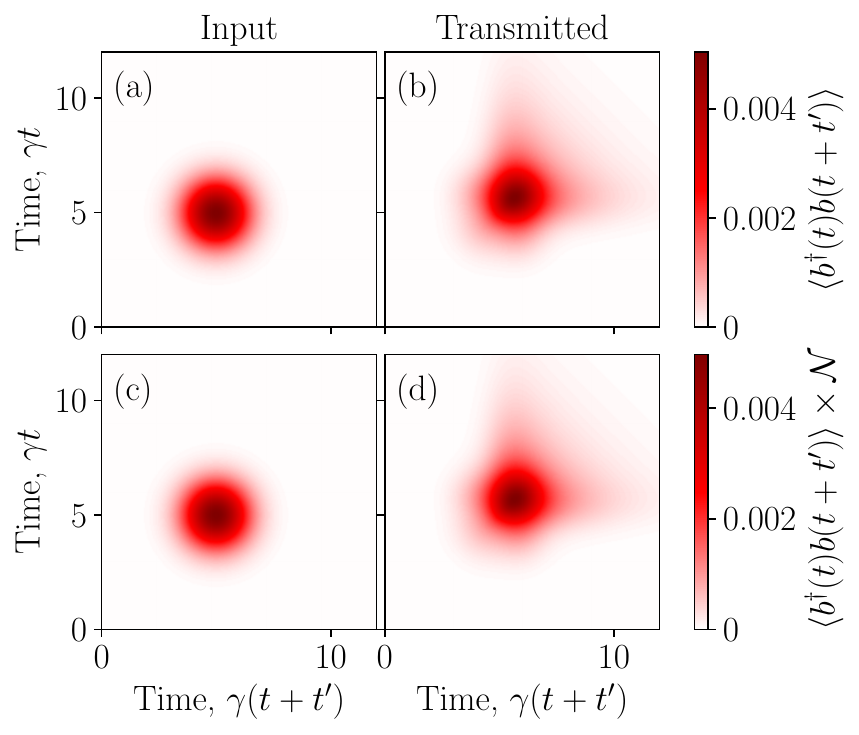}
    \caption{First-order correlation $\braket{b^{\dagger}(t) b(t+t')}$ for a Gaussian squeezed-state pulse with $\gamma \sigma_t=1$ and centered at $\gamma t_c=5$, going through a waveguide without any interaction in (a), and when interacting with a chiral TLS in (b).
    For comparison, panels (c,d) show the same first-order correlation of a pure 2-photon Fock Gaussian pulse,
    also with $\gamma \sigma_t=1$ and $\gamma t_c=5$. 
    All the results shown are scaled by $[\gamma]$.
    }
    \label{fig:g1}
\end{figure}

Figure~\ref{fig:g1} shows the first-order correlation  $\braket{b^\dagger(t) b(t+t')}$ for the
incident and transmitted fields of a squeezed
pulse as well as a 2-photon Fock state pulse. 
It is important to highlight that this correlation looks similar to the same correlation calculated for a pure two-photon Fock pulse, but scaled to the small squeezed parameter. 
Figure~\ref{fig:g1}(a) shows the input Gaussian quantum pulse shape as it travels the waveguide, while in Fig.~\ref{fig:g1}(b) we can observe the interaction with the TLS, with a small delay on the dynamics, which presents results similar to the Fock state equivalent, apart from a scaling factor [Fig.~\ref{fig:g1} (c,d)]. We also see two characteristic streaks which stem from
the TLS decay dynamics, as photons are re-emitted into the waveguide.

\begin{figure}[ht]
\centering
    \includegraphics[width=\columnwidth]{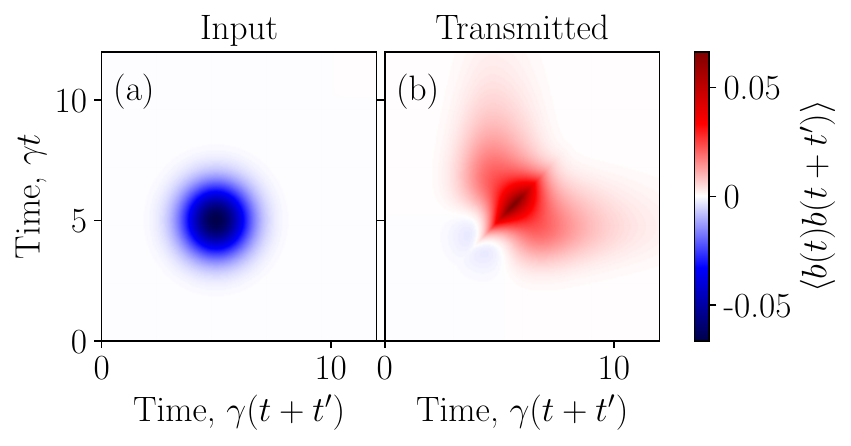}
    \caption{First-order correlation $\braket{b(t)b(t+t') }$ for a squeezed pulse without (a) and with (b) an interaction with a chiral TLS in a waveguide, input and transmitted, respectively. 
    We see a bird-like feature emerging that is similar to the 
    2-photon Fock state solution for 
    second-order correlation functions~\cite{le_jeannic_dynamical_2022,matias_2025,Nysteen2015}.
    Note the 2-photon Fock state solution in this case is identically zero. 
    Note that for a one-photon pulse or a linear pulse, this correlation function is formally zero, and it is also closely connected to squeezing.
    All the results shown are scaled by $[\gamma]$.
    }
    \label{fig:bb}
\end{figure}

Until now, we have observed correlations that are similar to the 2-photon Fock state counterpart. However, we are interested in studying observables that are unique to the squeezing solution. Thus, in Fig.~\ref{fig:bb}, we calculate $\braket{b(t+t') b(t)}$.

Figure~\ref{fig:bb} shows results 
for the correlation $\braket{b(t)b(t+t')}$
from a squeezed pulse, which produced results
that are now completely different from those 
of a 2-photon Fock state pulse. When working with Fock state pulses, this observable will always remain zero. However, with a squeezed pulse, we can observe finite values of $\braket{b(t) b(t+t')}$. First, in Fig.~\ref{fig:bb}(a), we see the Gaussian pulse shape again, but now there has been a sign flip in comparison to $\braket{b^\dagger(t) b(t+t')}$. In addition, its maximum value is an order of magnitude larger than $\braket{b^\dagger(t+t') b(t)}$. This not only means that there is in fact squeezing, but also that this squeezing correlation is dominant. Thus, even for weak pulses, such as is the case of $\zeta=0.1$, one can already observe significant nonlinear effects with these correlation functions.

Interestingly, in Fig.~\ref{fig:bb}(b), when we add a TLS interacting with our squeezed pulse, the main interaction flips signs again, and we observe a bird-like pattern that is close to the one that has been observed for second-order correlation results when working with 2-photon Fock state pulses~\cite{le_jeannic_dynamical_2022,matias_2025,Nysteen2015,regidor2026quantumdynamicsfewphotonpulsed}, although in this case, we have positive and negative values.

\begin{figure}[ht]
\centering
    \includegraphics[width=\columnwidth]{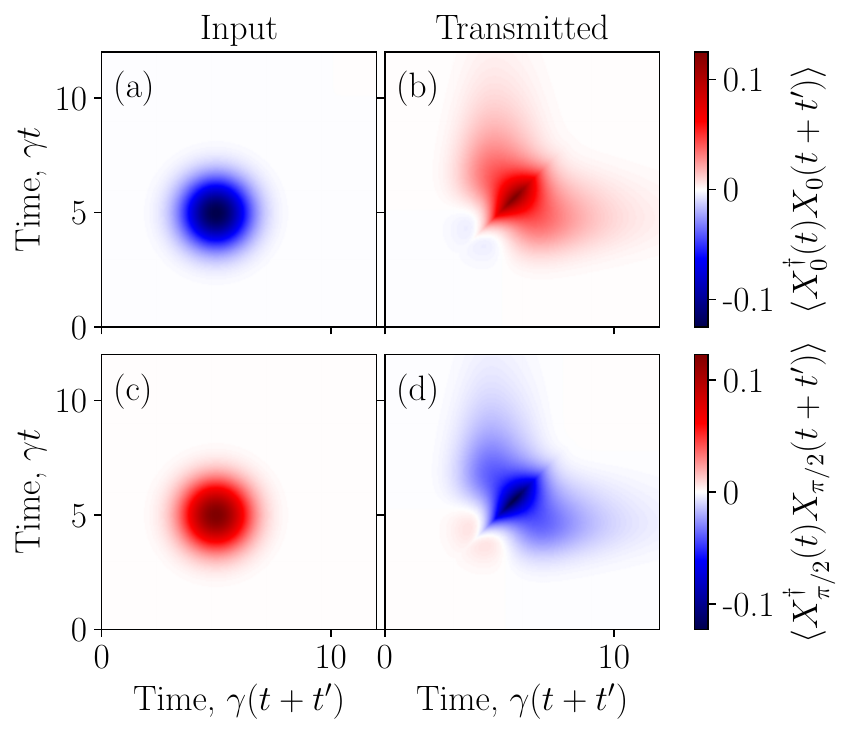}
    \caption{First-order correlation $\braket{X_\phi^{\dagger}(t) X_\phi(t+t')}$ for the case of $\phi=0$ (a,b), for a pulse without any interaction (input) in (a), and when it interacts with a chiral TLS (transmitted) in (b). In (c,d), similar results are shown for the case of $\phi=\pi/2$, for a pulse without any interaction in (c), and interacting with a chiral TLS in (d). All the results shown are scaled by $[\gamma]$.
    }
    \label{fig:X_dag_X}
\end{figure}

Figure~\ref{fig:X_dag_X} next presents correlations between quadrature observables $\braket{X^\dagger_\phi(t) X_\phi(t+t')}$, in Fig.~\ref{fig:X_dag_X}(a,b) for a zero quadrature angle ($\phi=0$), and in 
Fig.~\ref{fig:X_dag_X}(c,d) for a quadrature angle of $\phi=\pi/2$.  
If we first take a look at the case of $\phi=0$, we can observe how correlations differ from the ones shown in Fig.~\ref{fig:g1}. However, in the case of using Fock or coherent pulses, all the quadrature values related to $b^2$ would be zero, and these two observables would have similar dynamics (where $\braket{X^\dagger_0 X_0}$ would appear to be just double the values of $b^\dagger b$). However, these squeezed pulses are dominated by the term $b^2$, as we saw in Fig.~\ref{fig:bb} with values of an order of magnitude larger. Thus, $X^\dagger X$ follows a result that is closer to two times the results observed in Fig.~\ref{fig:bb}. 

Next, if we consider the case of $\phi=\pi/2$ [Fig.~\ref{fig:X_dag_X} (c,d)], where again we show both without and with the TLS interaction, we see similar dynamics to those in (a) and (b), respectively, but with a flip in the sign of the values in both cases.

We also see how the correlations reshape in all the cases in which the pulse interacts with the TLS, with correlation patterns that deviate from the original Gaussian shape of the input pulse.

\subsection{Second-order correlations}
\label{subsec:res_secondcor}

\begin{figure}[t]
\centering
     \includegraphics[width=\columnwidth]{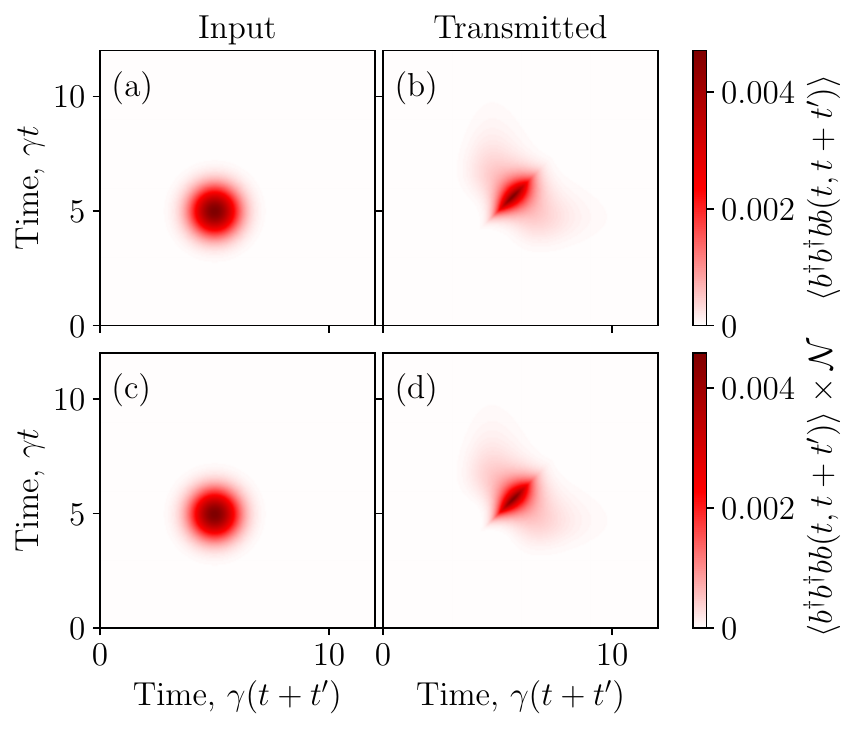}
    \caption{Second-order correlation function $\braket{b^{\dagger}(t) b^{\dagger}(t+t') b(t+t') b(t)}$ ($ \equiv \braket{b^{\dagger} b^{\dagger} b b(t,t+t')}$ in the figure for shorter notation), for a squeezed pulse without any interaction (input) in (a), and when it interacts with a chiral TLS (transmitted) in (b).
    For comparison, panels (c,d) show the same second-order correlation of a pure 2-photon Fock Gaussian pulse,
    also with $\gamma \sigma_t=1$ and $\gamma t_c=5$. All the results shown are scaled by $[\gamma^2]$.
    }
    \label{fig:g2}
\end{figure}

\begin{figure}[h]
\centering
     \includegraphics[width=\columnwidth]{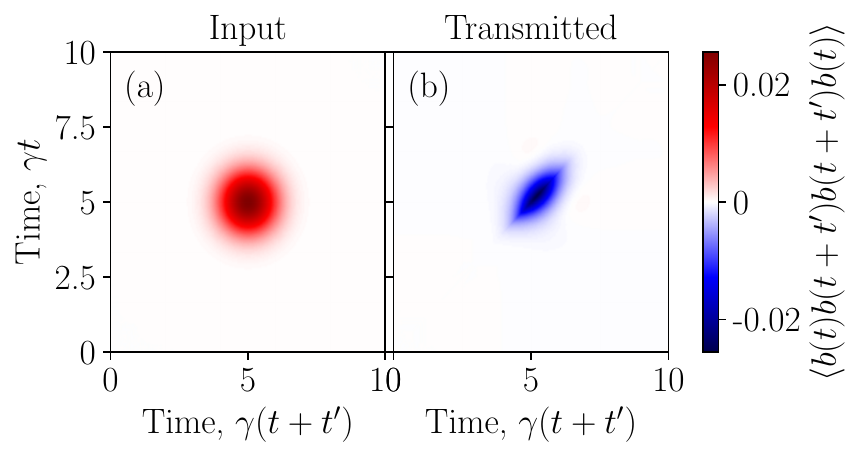}
     \caption{Second-order correlation $\braket{b(t) b(t+t') b(t+t') b(t)}$ for a squeezed pulse without any interaction (input) in (a), and when it interacts with a chiral TLS (transmitted) in (b). All the results shown are scaled by $[\gamma^2]$.
     }
    \label{fig:bbbb}
\end{figure}
\begin{figure*}[t]
\centering
     \includegraphics[width=0.99\textwidth]{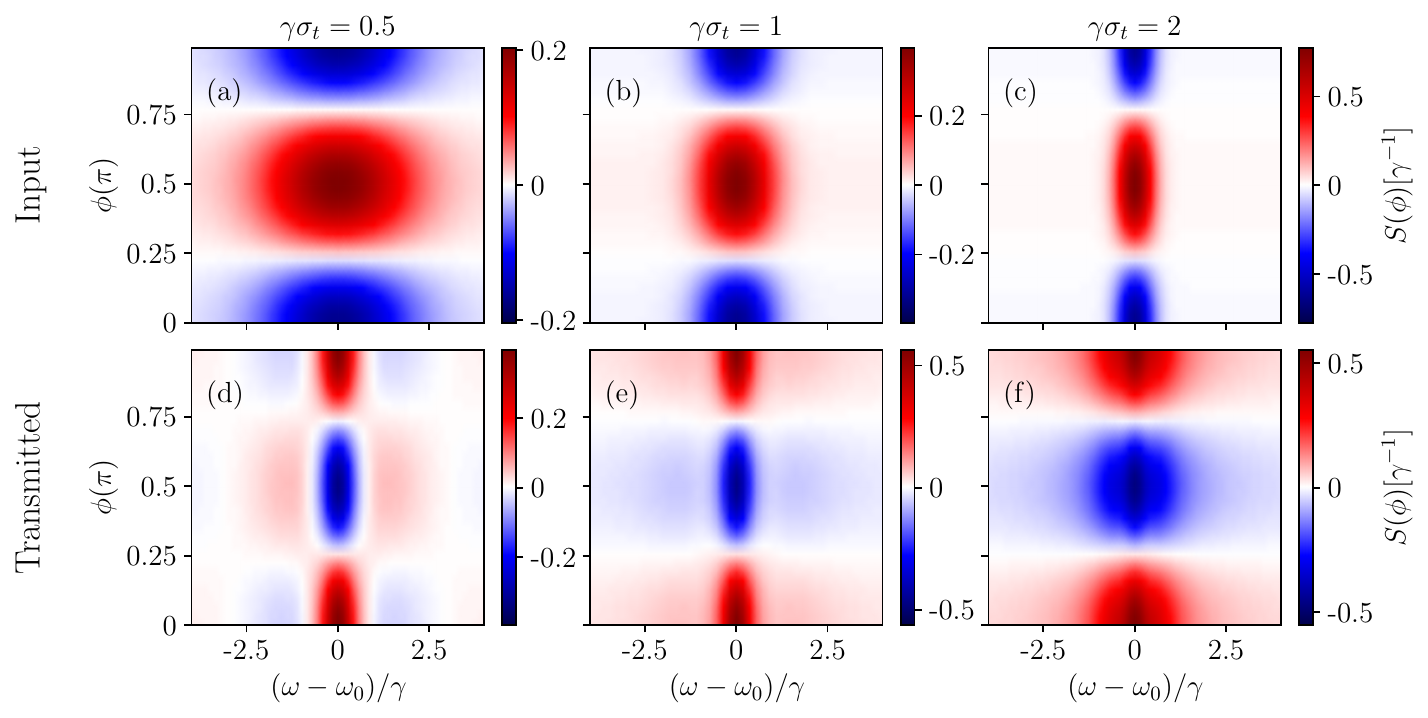}
     \caption{Squeezing spectra as a function of squeezing angle normalized to $\pi$, $\phi(\pi)$, of a squeezed pulse without any interaction (input) in (a,b,c), and when it interacts with a chiral TLS (transmitted) in (d,e,f) for three different pulse time lengths: $\gamma \sigma_t=0.5$ (a,d), $\gamma \sigma_t=1$ (b,e), and $\gamma \sigma_t=2$ (c,f).
     }
    \label{fig:S_sq}
\end{figure*}

In this subsection, we can consider the usual second-order correlation function $\braket{b^{\dagger}(t) b^{\dagger}(t+t') b(t+t') b(t)}$. 
This is a common experimentally measured
observable with weak coherent states and 
2-photon Fock state pulses \cite{le_jeannic_dynamical_2022}. 
This is shown in Fig.~\ref{fig:g2}(a) for a squeezed pulse without interaction, and in (b) when adding the TLS interaction. Once again, this observable $\braket{b^{\dagger}(t) b^{\dagger}(t+t') b(t+t') b(t)}$ presents results that are very similar to the ones observed with a 2-photon Fock state pulse, but with a smaller maximum value due to the squeezed pulse strength [Fig.~\ref{fig:g2} (c,d)].

However, we can also consider the following correlation: $\braket{b(t) b(t+t') b(t+t') b(t)}$, as shown in  Fig.~\ref{fig:bbbb}. Similar to the previous case with the first-order $b^2$ correlation, even though for Fock pulses $\braket{b(t) b(t+t') b(t+t') b(t)}$ is zero, we can observe finite values of this observable with our squeezed state. This is a remarkable result, since we are working with a weak pulse and this is a unique quantum feature from the 4-photon component.  

Surprisingly, we see that even for small values of $\zeta$, where the expected number of photons remains very low ($\braket{n} \approx 0.01$), we need to consider at least 4 photons before truncating the system in order to calculate this observable.

\subsection{Squeezing Spectra}
\label{subsec:res_spectra}

Finally, we study the squeezing spectra. In Figs.~\ref{fig:S_sq}(a,b,c), we see the case again first without any TLS interaction, i.e. the input pulse solution. We study this for three different pulse time lengths: $\gamma \sigma_t=0.5$ (a), $\gamma \sigma_t=1$ (b), and $\gamma \sigma_t=2$ (c), and observe how this is reflected in the spectra. The longer the time length is, the narrower the spectrum becomes.
Here, we can also clearly see the squeezing effects, showing negative spectral values that depend on the angle. 
These have the same qualitative features as squeezed wave packets that have been generated experimentally, e.g., by exciting nonlinear cavities or by driving quantum dots with a weak coherent field~\cite{PhysRevLett.125.170402,Patel2026}.

Then, in Figs.~\ref{fig:S_sq}(d,e,f), we study the interaction with a TLS for the same temporal pulse lengths. In these cases, the results become more complicated, following the nonlinear interaction with the TLS. First, we observe a clear flip of the sign for all the values of $\phi$, but we can also observe how the spectrum narrows, which is related to the nonlinear interaction between the squeezed pulse and the TLS. In this case, the squeezing spectrum narrows down and peaks at higher values due to the TLS interaction (and its decay rate). Thus, for the shortest pulse length in (d), we can still observe some values that have not flipped sign. However, when the temporal length becomes longer, and thus, the spectrum gets narrower, such as in (f), the flip of signs is total, and the central peak is sharper. Therefore, the temporal length of the squeezed pulse is going to determine how the squeezed spectrum looks after interacting with the TLS.

\section{Conclusions}
\label{sec:conclusions}

We have developed a powerful MPS framework to model few-photon squeezed vacuum pulses interacting with a TLS in a waveguide, where waveguide photons are treated exactly (at the system level, along with the TLS). By constructing the input pulse directly as a superposition of Fock-state components within the MPS formalism, our approach enables numerically exact time-domain simulations of pulsed squeezed states with arbitrary temporal envelopes. Moreover, this method naturally describes squeezing with finite bandwidth and provides direct access to both local observables and multi-time quantum correlation functions and squeezing spectra.

We have shown that, in the regime of weak squeezed pulses, several observables such as the emitter population, transmitted photon flux, and the conventional first- and second-order correlation functions are mainly dominated by the two-photon component of the squeezed state, showing similar dynamics to a pure two-photon Fock-state pulse after an appropriate normalization. This gives us a direct connection between the nonlinear response of squeezed pulses and few-photon nonlinearities of waveguide QED.

Furthermore, we have shown here observables that are unique to squeezed light, which have no counterpart for Fock-state pulses, such as $\braket{b(t)b(t+t^\prime)}$ and $\braket{b(t)b(t+t^\prime)b(t+t^\prime)b(t)}$, which
have finite values and interesting quantum correlations even when considering very weak pulses. Here, we have also shown that after interaction with the TLS, $\braket{b(t)b(t+t^\prime)}$ follows a characteristic bird-like feature, resembling results previously associated with second-order correlations of the scattering of 2-photon Fock pulses.
These results demonstrate that squeezed pulses contain rich nonlinear light-matter interactions captured by these correlation functions available even with low photon numbers.

Finally, we have investigated the squeezing spectra before and after the interaction with a TLS, demonstrating how the spectral response depends on both the pulse temporal length and the scattering process. 
The interaction with the TLS reshapes the squeezing spectrum by flipping its sign, and the bandwidth of the pulse reshapes the width of the spectrum. These results highlight how quantum nonlinearities depend on finite pulse duration, squeezing, and the interaction
with the TLS.

This MPS method is a powerful numerical technique for studying pulsed squeezed states in waveguide QED beyond the CW limit. Since it is not restricted to Markovian dynamics, it can be extended to more complex non-Markovian architectures, including delayed coherent feedback and multi-emitter waveguide-QED systems, allowing one to explore new regimes of nonlinear quantum optics that are inaccessible with existing analytical techniques.

Our general results use parameters (in scaled units) that are consistent with several experimental platforms, including semiconductor systems (photonics regime) and circuit-QED systems (microwave regime), and thus our predictions should be accessible using current-day and emerging quantum technologies.

\acknowledgements
This work was supported by the Natural Sciences and Engineering Research Council of Canada (NSERC) (Discovery Grant and Quantum Alliance); the Canadian Foundation for Innovation (CFI); and Queen's University, Canada.
We thank Marc Dignam for useful discussions.

\section*{DATA AVAILABILITY}

All the MPS calculations performed in this paper were carried out with the open-source Python package QwaveMPS~\cite{2602.15826}.

\appendix

\section{Benchmarking the population dynamics}
\label{appendix}

To corroborate the accuracy of our MPS formulism,
in this Appendix, we
compare and benchmark some of our MPS population results against an alternative method described in Ref.~\cite{PhysRevA.105.023721}, where they derived squeezed-wave-packet master equations,
that give a solution for the
populations in terms of an infinite set of coupled equations, which requires one to introduce
a numerical cut-off (for the hierarchy).

\begin{figure}[h]
\centering
    \includegraphics[width=\columnwidth]{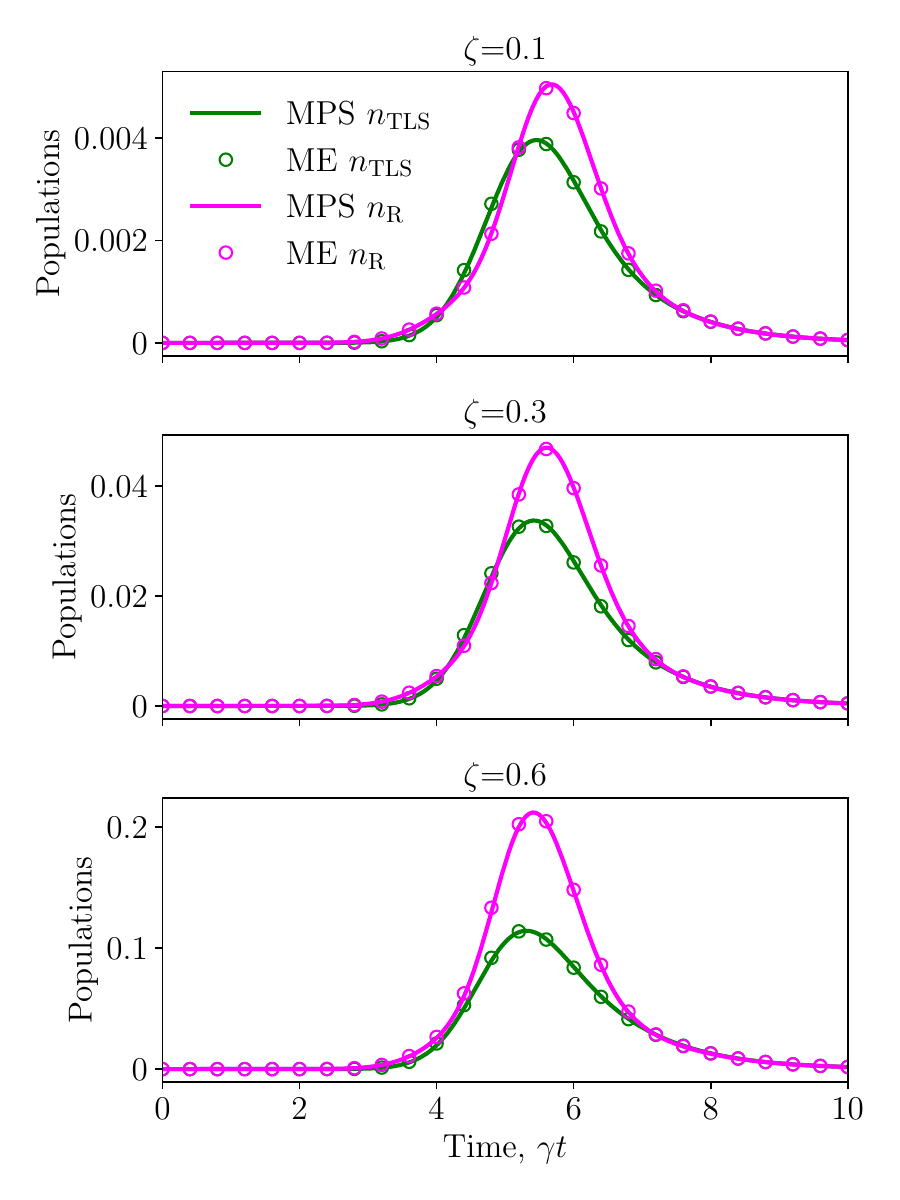}
    \caption{
    Population dynamics for a squeezed vacuum Gaussian pulse with $\gamma \sigma_t=1$ and centered at $\gamma t_c=5$, interacting with a chiral TLS, with three different squeezing parameters: $\zeta=0.1$, $\zeta=0.3$ and $\zeta=0.6$. Two-level system population dynamics $n_{\rm TLS}$ is shown in green, and right transmitted flux $n_{\rm R}$ in magenta. Solid lines represent our MPS calculations, and circles show the squeezed-wave-packet master equation results from Ref.~\cite{PhysRevA.105.023721}.
    }
    \label{fig:benchmark}
\end{figure}

Specifically, we use their equation (5.5)~\cite{PhysRevA.105.023721} (we use ``equation'' here to reference equations from another paper) to calculate the TLS population, shown in Fig.~\ref{fig:benchmark} in green circles, and compare it with our MPS counterpart, shown in green solid lines;  we observe excellent agreement. In addition, we also compared the photon flux by solving their equation~(5.12) (Fig.~\ref{fig:benchmark} in magenta circles) and the photon flux calculated using our approach (magenta solid lines) and again see excellent agreement between methods. We benchmark these observables for three different squeezed parameters ($\zeta=0.1$, $\zeta=0.3$ and $\zeta=0.6$) to ensure agreement in all cases. For the MPS calculations, we have limited the photon subspaces to 4,6, and 8, respectively, and used Eq.~\eqref{eq:trunc} to control their convergence.




It is important to note here that, although the method studied in Ref.~\cite{PhysRevA.105.023721} is useful for benchmarking these population observables, our MPS approach is 
very general, giving us full access to multi-time photon correlations in a straightforward way.
In addition, we can easily add 
other features, such as a symmetrically coupled TLS where two channels (left and right) are required, non-Markovian dynamics with time-delayed feedback, and multi-emitter effects (with also long
spatial separations). 

\clearpage

\bibliographystyle{quantum}
\bibliography{biblio}

\end{document}